\documentclass[pre, twocolumn, amsmath, amssymb]{revtex4}

\usepackage[dvips]{graphicx}

\let\footnote\savefootnote
\begin{document}

\title[Fractal time random walk, laser cooling and renewal processes]
{Fractal time random walk and subrecoil laser cooling \\
considered as renewal processes with infinite mean waiting times}


\author{F. Bardou}
\affiliation{IPCMS, CNRS and Universit\'e Louis Pasteur
23 rue du Loess, BP 43, F-67034 Strasbourg Cedex 2, France}
\email{bardou@ipcms.u-strasbg.fr}

\begin{abstract}
There exist important stochastic physical processes involving {\it infinite}
mean waiting times. The mean divergence has dramatic consequences 
on the process dynamics. Fractal time random walks, a diffusion process, 
and subrecoil laser cooling, a concentration process, are
two such processes that look qualitatively dissimilar. Yet,
a unifying treatment of these two processes, which is the topic 
of this pedagogic paper, can be developed by combining
renewal theory with the generalized central limit theorem. 
This approach enables to derive without technical difficulties 
the key physical properties and it emphasizes the role of the behaviour
of sums with infinite means.
\end{abstract}


\maketitle

To appear in: 
{\small \it Proceedings of Cargese Summer School on ``Chaotic Dynamics 
and Transport in Classical and Quantum Systems'', August 18-30 (2003).}


\section*{Introduction}
The fractal time random walk \cite{Shl1974,SSB1991} 
has been developed in the 1970's to explain 
anomalous transport of charge carriers in disordered solids. It describes 
a process in which particles jump from trap to trap
as a result of thermal activation with a 
very broad (infinite mean) distribution of trapping times. It results in 
an unusual time-dependence of the position distribution which broadens
while the peak remains at the origin. The
method of choice to study the fractal time random walk is the continuous
time random walk technique.

Subrecoil laser cooling \cite{AAK1988,BBA2002} 
has been developed in the 1990's as a way to reduce
the thermal momentum spread of atomic gases thanks to momentum exchanges
between atoms and laser photons. It is a process in which, 
as a result of photon scattering, atoms jump
from a momentum to another one  
with a very broad distribution of waiting times between two scattering events.
It results in an unusual time dependence of the momentum distribution which 
narrows without fundamental limits hence giving access to temperatures in 
the nanokelvin range. The method of choice to study subrecoil laser cooling
is renewal theory \cite{BaB2000}.

Fractal time random walks and subrecoil cooling seem at first sight very 
dissimilar. The first mechanism generates a broader and broader distribution,
while the second generates a narrower and narrower distribution. 
Nevertheless, inspection
of the theories of both phenomena reveals strong similarities: 
the continuous time
random walk and the renewal theory are two closely related
ways to tackle related stochastic processes. 
Physically, the two mechanisms share
a common core, a jump process with a broad distribution of waiting times.

\smallskip
The aim of this pedagogic paper is to bridge the gap between fractal time
random
walk and subrecoil laser cooling. We show that the essential results of the  
two theories can be obtained nearly without calculation 
by combining the simple probabilistic
reasoning underlying renewal theory and the generalized central limit
theorem applying to broad distributions. This provides 
more direct derivations than in earlier approaches, at least for
the basic cases considered here.

In the first part, we describe the microscopic stochastic mechanisms 
at work in the fractal time random walk and in subrecoil cooling and 
relate them to renewal processes. In the second part, 
we explain elementary properties of renewal theory and derive asymptotic
results using the generalized central limit theorem and L\'evy stable 
distributions. In the third part, we draw the consequences for the fractal
time random walk and subrecoil cooling. The fourth part contains 
bibliographical notes.

\section{Fractal time random walk and subrecoil cooling: microscopic mechanisms}
\label{s1}

\subsection{Fractal time random walk}
\label{s1.1}
The notion of fractal time random walk emerged from the observation of unusual
time dependences in photoconductivity transient currents flowing through 
amorphous samples. It can be schematized in the following way.

\begin{figure}[ht]
\includegraphics[scale=0.55,angle=0]{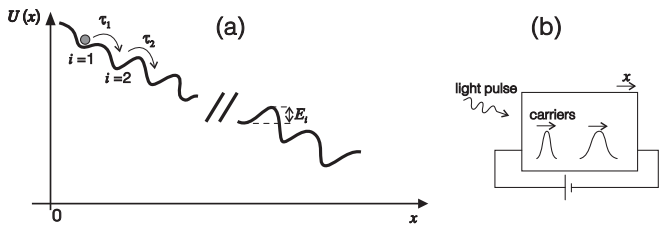}
\caption{Biased random walks in a disordered system 
(a) {\em Arrhenius cascade}. Each carrier is put in random potential 
with a global tilt. It undergoes jumps over barriers of random heights
$E_i$ from a metastable well to the next one on the right thanks 
to thermal activation. (b) {\em Photoconductivity setup}. At time $t=0$,
a light pulse creates carriers in the immediate vicinity of the left 
electrode of an amorphous sample. The carriers of one sign then move 
through the sample thanks to an applied electric field. (The carriers of
the opposite sign are immediately absorbed by the left electrode.)}
\label{fig1_1}
\end{figure}

Consider first a one dimensional situation called the Arrhenius cascade 
\cite{Bar1999} in which the charge carriers are placed in a random potential
with many local wells and barriers and can jump from one well to another 
one thanks to thermal activation (Fig.~\ref{fig1_1}a). The Arrhenius cascade
potential presents two features: a global tilt representing the effect 
of the electric field on the carriers and local random oscillations 
creating metastable traps separated by barriers representing the disorder
created by the amorphous material. Thus the potential seen by the carriers
is a kind of random washboard with a discrete number of metastables states.

The mean lifetime of state $i$, {\it i.e.}, the mean waiting time before 
the occurrence of a thermal jump, is given by the Arrhenius law:
\begin{equation}
\bar \tau_i = \tau_0 e^{E_i/kT}
\end{equation}
where $\tau_0$ is a time scale, $E_i$ is the height of the energy barrier 
separating state $i$ from state $i+1$, $k$ is the Boltzmann constant
and $T$ is the temperature. The potential global tilt is assumed to be 
large enough to neglect backward jumps from $i$ to $i-1$. 
The random walk we consider is thus completely biased. 
The time spent between the metastable states is neglected.
For a given barrier height $E_i$, the lifetime distribution $\psi(\tau|E_i)$
is exponential with mean $\bar\tau_i$:
\begin{equation}
\psi(\tau|E_i) = \frac{1}{\bar \tau_i} \ e^{-\tau/\bar\tau_i}.
\label{e1.1.2}
\end{equation}

\medskip
In the photoconductivity experiments (Fig.~\ref{fig1_1}b), 
a light pulse creates at time $t=0$
carriers localized near the surface of the sample. The carriers 
then move through the disordered sample thanks to an applied electric field.
Thus, this situation can be modelled by a large number of Arrhenius cascade 
in parallel, each electron path being associated to one cascade. 

One may (wrongly) expect that the transient current flowing through the sample
is quasi-constant at the beginning, 
while the bunch of carriers propagates through
the sample, before decreasing rapidly to zero when the carriers
leave the sample after reaching the end electrode.
But what is observed is quite different. The current decreases as a power law
$\sim 1/t^{1-\alpha}$ while the carriers are still in the sample, then
as $\sim 1/t^{1+\alpha}$ when some carriers start leaving the sample.
For simplicity, we assume here that the sample is semi-infinite so that 
the carriers never leave the sample.

The explanation of this anomalous behaviour will be shown to be related 
to the distribution of lifetimes $\tau_i$. The randomness of the 
$\tau_i$'s results from the combination of the exponential statistics 
of jump times for a given barrier height $E_i$ (eq.~(\ref{e1.1.2})) 
with the barrier height statistics conveniently described by an exponential
distribution $P(E_i)$,
\begin{equation}
P(E_i) = \frac{1}{E_0} \ e^{-E_i/E_0} \quad \mathrm{for} \quad E_i \geq 0,
\end{equation}
where $E_0$ is an energy scale related to the sample disorder.

The waiting time distribution $\psi(\tau)$ is then 
\begin{equation}
\psi(\tau) = \int_0^\infty \mathrm{d} E_i \ P(E_i) \psi(\tau|E_i) = 
\alpha \gamma(1+\alpha,\tau/\tau_0) \ \frac{\tau_0^\alpha}{\tau^{1+\alpha}}
\end{equation}
where $\gamma(\alpha',x) = \int_0^x e^{-u} u^{\alpha'-1} \mathrm{d} u$
is the incomplete gamma function and
\begin{equation}
\alpha = \frac{kT}{E_0}.
\end{equation}
At long times, $\psi(\tau)$ tends to a power law, hence the term 
``fractal time random walk":
\begin{equation}
\psi(\tau) \simeq \alpha^2 \Gamma(\alpha)
\frac{\tau_0^\alpha}{\tau^{1+\alpha}},
\end{equation}
with $\Gamma(\alpha) = \int_0^\infty u^{\alpha-1} e^{-u} \ \mathrm{d}u$.

If $\alpha \leq 1$, states $i$ have an infinite mean lifetime 
$\langle \tau \rangle
= \int_0^\infty \tau \psi(\tau) \ \mathrm{d}\tau$. 
However, they are unstable since they all ultimately decay to the 
next state $(i+1)$. Usually, unstable states have a well defined and finite 
mean lifetime. Here, the somewhat paradoxical presence of {\it unstable states
with infinite mean lifetimes} is at the origin of the striking properties
of the fractal time random walk.

\subsection{Subrecoil laser cooling}
\label{s1.2}

\begin{figure}[ht]
\includegraphics[scale=0.75,angle=0]{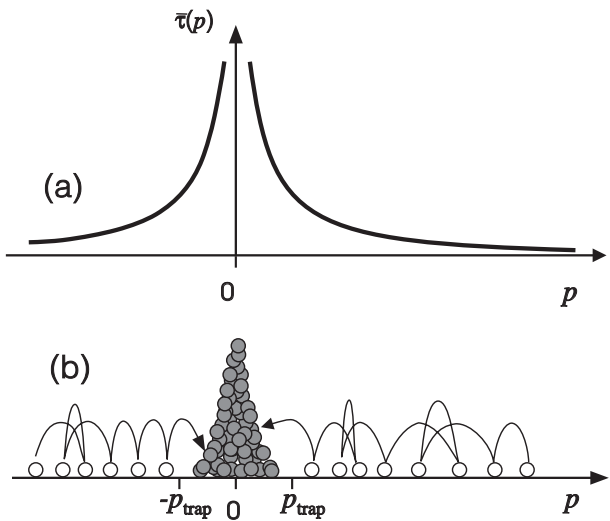}
\caption{Subrecoil laser cooling (a) The mean sojourn time at momentum $p$
$\bar\tau(p)$ becomes very large for small atomic momenta.
(b) Photon scattering creates a momentum random walk with an accumulation
in the vicinity of $p=0$ due to the momentum dependence of the 
mean sojourn time $\bar\tau(p)$.}
\label{fig1_2}
\end{figure}

Laser cooling of atomic gases consists in reducing the momentum spread 
of atoms thanks to momentum exchanges between atoms and photons. 
{\it Subrecoil} laser cooling consists in reducing the momentum spread
to less than a single photon momentum, denoted $\hbar k$. 
This paradoxical goal is achieved
by introducing a momentum dependence in the photon scattering rate 
(see Fig.~\ref{fig1_2}a) so that
it decreases strongly or even vanishes in the vicinity of $p=0$, where
$p$ denotes the atomic momentum, taken in one dimension for simplicity.

The mechanism of subrecoil cooling is explained in Fig.~\ref{fig1_2}.
Any time a photon is absorbed and spontaneously reemitted by an atom,
the atomic momentum undergoes a momentum kick on the order of $\hbar k$,
which has a random component because spontaneous emission occurs in a random
direction. Thus, the repetition of absorption-spontaneous emission cycles
generates for the atom a momentum random walk (see Fig.~\ref{fig1_2}b), 
with momentum dependent 
waiting times $\tau$ between two kicks. When an atom reaches by chance
the vicinity of $p=0$, it tends to stay there a long time. This enables
to accumulate atoms at small momenta, {\it i.e.}, to cool. 

For a quantitative treatment, we introduce $\bar \tau(p)$, the mean sojourn
time at momentum $p$ (also the mean waiting time between two spontaneous 
photons for an atom at momentum $p$). For a given $p$, the sojourn time
at momentum $p$, {\it i.e.}, the distribution $\psi(\tau|p)$ 
of sojourn times at momentum $p$ is
\begin{equation}
\psi(\tau|p) = \frac{1}{\bar \tau(p)} \ e^{-\tau/\bar\tau(p)}.
\label{e1.2.1}
\end{equation}
We need to characterize the 
distribution of ``landing" momenta $\pi(p)$ after a spontaneous emission.
Under favourable but often realistic assumptions, 
atoms spend most of the time around the origin in the interval 
$[-p_{\mathrm{trap}},+p_{\mathrm{trap}}]$, because they diffuse fast 
outside this interval and thus come back to it rapidly after leaving it. 
If $p_{\mathrm{trap}}< \hbar k$, then after a spontaneous emission, the 
distribution of atomic momenta can be considered as uniform:
\begin{equation}
\pi(p) = \frac{1}{2 p_{\mathrm{trap}}}.
\label{e1.2.2}
\end{equation}
The distribution of sojourn times after a spontaneous emission is thus
\begin{equation}
\psi(\tau) = \int_{-p_{\mathrm{trap}}}^{p_{\mathrm{trap}}} \mathrm{d} p \
\pi(p) \psi(\tau|p).
\end{equation}
We consider the physically relevant case of power law $\bar \tau(p)$,
\begin{equation}
\bar \tau(p) = \frac{\tau_0 p_0^\beta}{|p|^\beta},
\label{e1.2.4}
\end{equation}
where $\beta >0$, and  $\tau_0$ and $p_0$ are time and momentum scales, 
respectively. Then, one finds, just as in the fractal time random walk,
a waiting time distribution with a power law tail:
\begin{equation}
\psi(\tau) = \frac{\alpha p_0}{p_{\mathrm{trap}}} \ 
	\gamma\left[1+\alpha, 
			\left( \frac{p_{\mathrm{trap}}}{p_0}\right)^{1/\alpha} 
			\frac{\tau}{\tau_0} \right]
		\frac{\tau_0^\alpha}{\tau^{1+\alpha}}
\mathrel{\mathop{\kern 0pt\longrightarrow}\limits_{\tau\to \infty}}
\alpha^2 \Gamma(\alpha) \frac{p_0}{p_{\mathrm{trap}}} 
\frac{\tau_0^\alpha}{\tau^{1+\alpha}},
\label{e1.2.5}
\end{equation}
where 
\begin{equation}
\alpha = \frac{1}{\beta}.
\end{equation}
If $\beta<1$, the mean waiting time is finite and simple integration gives
\begin{equation}
\langle \tau \rangle = \frac{1}{1-\beta} 
\left( \frac{p_0}{p_{\mathrm{trap}}}\right)^\beta \tau_0.
\label{e1.2.7}
\end{equation}
If $\beta \geq 1$, on the contrary, the mean waiting time is infinite. This 
divergence of the mean has dramatic (and positive in terms of cooling) 
consequences (see \S\ref{s3.2}).

\subsection{Connection with renewal theory}
\label{s1.3}

\begin{figure}[ht]
\includegraphics[scale=0.80,angle=0]{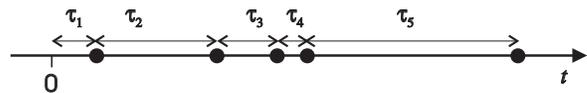}
\caption{Renewal processes. The system undergoes a sequence
of events (jumps from trap to trap, momentum kicks ...) at random
times separated by waiting times $\tau_1$, $\tau_2$, ...}
\label{fig1_3}
\end{figure}

Renewal processes are stochastic process in which a system 
undergoes a sequence of events (denoted by $\bullet$ in Fig.~\ref{fig1_3}) 
separated by independent random ``waiting times''
$\tau_1$, $\tau_2$, ... 
The term ``renewal process'' comes from engineering. Assume that,
at time $t=0$, one installs a machine in a factory. When, after being
operated for a random lifetime $\tau_1$, the machine breaks down,
it has to be replaced by a new one,
which will work till it breaks down at $\tau_1 + \tau_2$ and
has to be replaced ...  If, instead of a single machine, one has installed
a large number of identical machines, then, to decide how many
replacement machines must be stored at a given time, one needs to
know the replacement rate, which we call hereafter the renewal density.

To understand the statistical properties of renewal processes, various
quantities are introduced. The most detailed information
is provided by the {\it distribution of the number of renewals}, 
$f_t(r)$, {\it i.e.}, 
the probability distribution for the system to undergo $r$ events in time $t$.
One also introduces derived quantities, the {\it mean number of renewals} 
at time $t$, $\langle r \rangle_t$, and the mean renewal rate at time $t$,
denoted $R(t)$ and called the {\it renewal density}. 
Mathematical expressions for
these three quantities will be given in \S\ref{s2.1}. Here we show the role 
they play in fractal time random walk and in subrecoil laser cooling.

\medskip
In the {\it biased} (fractal time or non fractal time) {\it random walk}, 
the discretized positions 
$n$ at time $t$ correspond directly to the number of jumps performed between
time $0$ and $t$. Hence the position distribution $\rho(n,t)$ in the fractal
time random walk is the renewal number distribution:
\begin{equation}
\rho(n,t) = f_t(n).
\end{equation}

The mean position of the carriers is $\langle r \rangle_t$.
The current $i(t)$ measured in photoconductivity experiments before the carriers
get out of the sample is proportional to the mean carrier velocity 
$\mathrm{d}\langle r \rangle_t / \mathrm{d} t$, which is the renewal density 
(see \S\ref{s2.1}). Thus, one has 
\begin{equation}
i(t) \propto R(t).
\label{e1.3.2}
\end{equation}

\medskip
In {\it subrecoil cooling}, the momentum distribution $\rho(p,t)$ 
can be written in the following form:
\begin{equation}
\rho(p,t) = \pi(p) \int_0^t \mathrm{d} t_l \ R(t_l) P_s(t-t_l|p),
\label{e1.3.3}
\end{equation}
where $t_l$ is the time of the last jump occurring between $0$ and $t$,
$R(t_l) \mathrm{d} t_l$ is the probability that a jump occurs during 
the interval $[t_l, t_l+\mathrm{d} t_l)$, $\pi(p)$ is the uniform 
probability distribution for a jumping atom to land at momentum $p$
and $P_s(t-t_l|p)$ is the survival probability for an atom landing at 
momentum $p$ at time $t_l$ to stay there till at least time $t$.
Using eq.~(\ref{e1.2.1}), one has trivially
\begin{equation}
P_s(t-t_l|p) = \int_{t-t_l}^\infty \mathrm{d} \tau \ \psi(\tau|p) 
= e^{-(t-t_l)/\bar\tau(p)}.
\label{e1.3.4}
\end{equation}
The non trivial physical information is contained in the renewal density
$R(t)$.

The height $h(t)$ of the momentum distribution peak,
\begin{equation}
h(t) = \rho(p=0,t),
\end{equation}
is proportional to $\langle r \rangle_t$, the mean number of jumps
between $0$ and $t$. Indeed, for any jump, there is a probability 
$\pi(p) 2\mathrm{d}p = \mathrm{d}p / p_\mathrm{trap}$ to fall in the vicinity
$[-\mathrm{d}p, \mathrm{d}p]$ of the origin and to stay there indefinitely
since states in $[-\mathrm{d}p, \mathrm{d}p]$ have arbitrarily long lifetimes
in the limit $\mathrm{d}p \to 0$ 
($\bar\tau(p) \mathrel{\mathop{\kern 0pt\longrightarrow}\limits_{p\to 0}} 
\infty$, see eq.~(\ref{e1.2.4})). Thus the height writes
\begin{equation}
h(t) = \frac{\langle r \rangle_t}{2 p_\mathrm{trap}}.
\label{e1.3.6}
\end{equation}

\section{Renewal theory and L\'evy stable laws}
\label{s2}

\subsection{General formulae}
\label{s2.1}

The number of renewals $r_t$ in a time $t$ is defined as the number of jumps 
having occurred before time $t$. It satisfies
\begin{equation}
S_{r_t} \leq t < S_{r_t+1}
\end{equation}
where $S_{r_t} = \sum_{i=1}^{r_t} \tau_i$ is the sum of the first $r_t$ waiting
times. The relationship between the renewal number distribution $f_t(r)$
and the waiting time distribution $\psi(\tau)$ can be obtained 
from the following simple reasoning.

Note first that the distribution, denoted $\psi^{r*}(S_r)$, of the sum $S_r$
of $r$ independent identically distributed waiting times is the 
$r^{\mathrm{th}}$ convolution product of  $\psi(\tau)$ with itself.  
Moreover, from the definition of the number $r_t$ of renewals, one has obviously
\begin{equation}
\mathrm{Pr}(r_t<r) = \mathrm{Pr}(S_r > t) = \left[ 1 - \Psi^{r*}(t)\right],
\end{equation}
where $\Psi^{r*}(S_r) = \int_0^{S_r} \psi^{r*}(u) \ \mathrm{d} u$
denotes the distribution function of $S_r$ (in spite of its notation, 
$\Psi^{r*}(S_r)$ is {\it not} the $r^{\mathrm{th}}$ convolution product
of the waiting time distribution function
$\Psi(\tau) = \int_0^\tau \psi(u) \ \mathrm{d} u$).
The probability distribution $f_t(r)$ of the number of renewals at time $t$
is thus finally
\begin{equation}
f_t(r_t=r) = \mathrm{Pr}(r_t<r+1) -\mathrm{Pr}(r_t<r) = \Psi^{r*}(t) - 
\Psi^{(r+1)*}(t).
\label{e2.1.3}
\end{equation}
This expression relates the distribution of a discrete random 
variable, $r$, to the distribution fonctions of continuous random 
variables $t$. Important quantities derived from the renewal number distribution
$f_t(r)$ are the mean number of renewals at time $t$:
\begin{equation}
\langle r \rangle_t = \sum_{r=0}^{\infty} r f_t(r)
\end{equation}
and the renewal density, {\it i.e.}, the mean number of renewals per
unit time%
\footnote{The renewal density is a rate of events. It has the same dimension
as a probability distribution of times, 1/time, but it is {\it not}
a probability density. Thus, its integral $\int_0^\infty \mathrm{d}t \ R(t)$,
representing the average total number of events occurring between
$t=0$ and $t=\infty$ does not have to be normalized. It is actually
often infinite.}%
:
\begin{equation}
R(t) = \frac{\mathrm{d}\langle r \rangle_t}{\mathrm{d}t}.
\label{e2.1.5}
\end{equation}

Usual theoretical treatments of renewal problems are based on Laplace 
transforms of the waiting time distribution that are indeed well suited
to handle the convolution products $\psi^{r*}(t)$. Here, we prefer a different
approach based only on the {\it generalized central limit theorem}. 
This approach requires nearly no calculation. Moreover, it stresses 
the role of the behaviour of sums of many random variables.

Indeed, most useful distributions tend, under repeated convolution, 
to a L\'evy stable law given by the generalized central limit theorem.
Thus we can obtain from eq. (\ref{e2.1.3}) analytical
expressions for $f_t(r)$ and for related quantities in the limit of large $r$
and hence large $t$. Depending on the finiteness of the first two moments
of the waiting time distribution $\psi(\tau)$, three cases can be distinguished.
For brevity, we treat only the two most striking cases: first
and second moment both finite in \S\ref{s2.2}; first
and second moment both infinite in \S\ref{s2.3}.

\subsection{Case of waiting time distributions with finite first and
second moment}
\label{s2.2}

When both the mean $\mu = \langle \tau \rangle$ and the variance $\sigma^2$
of the waiting time are finite, the (usual) central limit theorem 
applies and $\psi^{r*}(t)$ tends to a Gaussian distribution
\begin{equation}
\psi^{r*}(t) \mathrel{\mathop{\kern 0pt\longrightarrow}\limits_{r\to \infty}} \frac{1}{\sqrt{2 \pi} \sigma_r} \exp 
\left[ -\frac{(t-\mu_r)^2}{2 \sigma_r^2} \right]
\end{equation}
with mean $\mu_r = r\mu$ and variance $\sigma_r^2 = r\sigma^2$.
Thus, after changing variables in eq. (\ref{e2.1.3}), one has
\begin{equation}
f_t(r) \mathrel{\mathop{\kern 0pt\longrightarrow}\limits_{r\to \infty}}
\int_{(t-\mu_{r+1})/\sigma_{r+1}}^{(t-\mu_r)/\sigma_r} 
\frac{1}{\sqrt{2\pi}} \ e^{-u^2/2} \ \mathrm{d}u.
\end{equation}
Fix $(t-\mu_r)/\sigma_r$ and take $t\to \infty$. One expects intuitively
that the number of renewals is approximately given by $t/\mu \gg 1$ at large
$t$ (this is validated a posteriori by eq.~(\ref{e2.2.4})). 
Thus $(t-\mu_r)/\sigma_r
\simeq (t-\mu r)/ (\sigma \sqrt{t/\mu})$ and
$(t-\mu_{r+1})/\sigma_{r+1} \simeq (t-\mu r)/ (\sigma \sqrt{t/\mu}) 
- \sqrt{\frac{\mu^3}{\sigma^2 t}}$.
This leads for the number of renewals to a Gaussian distribution with mean 
$t/\mu$ and variance $\sigma^2t/\mu^3$:
\begin{equation}
\begin{array}{|c|}
\hline
f_t(r) \mathrel{\mathop{\kern 0pt\longrightarrow}\limits_{t\to \infty}}
 \frac{1}{\sqrt{2\pi \sigma^2 t/\mu^3}} \exp \left[ - 
 \frac{(r-t/\mu)^2}{2 \sigma^2 t / \mu^3}\right] \ . 
 \\
\hline
\end{array}
\label{e2.2.3}
\end{equation}
Although well known in renewal theory, this result is non trivial: 
the distribution of the sums of $r$ ($\to \infty$) terms 
and of the number of renewals at large times are found to have the same 
(Gaussian) shape. This 
is grossly violated with infinite mean waiting times (see \S\ref{s2.3}).

As direct consequences of eq.~(\ref{e2.2.3}), the mean number of renewals 
$\langle r \rangle_t$ tends to 
$t/\mu$ at large times:
\begin{equation}
\langle r \rangle_t 
\mathrel{\mathop{\kern 0pt\longrightarrow}\limits_{t \to \infty}}
\frac{t}{\mu} 
\label{e2.2.4}
\end{equation}
and, using eq.~(\ref{e2.1.5}), the renewal density tends 
to the reciprocal of the mean waiting time:
\begin{equation}
R(t) 
\mathrel{\mathop{\kern 0pt\longrightarrow}\limits_{t \to \infty}}
\frac{1}{\mu} \ ,
\label{e2.2.5}
\end{equation}
in agreement with intuition.

Physical comments and an example of a renewal process with finite 
mean waiting time will be presented in \S\ref{s3.1}.

\subsection{Case of waiting time distributions with infinite first and
second moment}
\label{s2.3}

We consider now the case of waiting time distributions with infinite first 
two moments, focusing on the canonical example of distributions with
(Pareto) power law tails of index $\alpha$,
\begin{equation}
\psi(\tau) \mathrel{\mathop{\kern 0pt\longrightarrow}\limits_{\tau \to \infty}}
\frac{\alpha \tau_0^\alpha}{\tau^{1+\alpha}},
\label{e2.3.1}
\end{equation}
with $0<\alpha<1$ (slightly more general cases can be handled using the theory
of regular variations). The 
distributions $\psi^{r*}$ of the sums $S_r$ no longer tends to Gaussians 
but, according to the generalized central limit theorem, to L\'evy stable
laws:
\begin{equation}
\psi^{r*}(t) 
\mathrel{\mathop{\kern 0pt\longrightarrow}\limits_{r \to \infty}}
\frac{1}{r^{1/\alpha}} \ L_{\alpha,B} \left(\frac{t}{r^{1/\alpha}}\right)
\label{e2.3.2}
\end{equation}
where $L_{\alpha,B}(u)$ is a one-sided ($u\geq 0$) L\'evy stable law 
given by its Laplace transform,
\begin{equation}
\int_0^\infty e^{-su} L_{\alpha,B}(u) \ \mathrm{d}u = e^{-B s^\alpha},
\label{e2.3.3}
\end{equation}
and $B$ is a scale parameter given by
\begin{equation}
B = \Gamma(1- \alpha)\tau_0^\alpha.
\label{e2.3.4}
\end{equation}
After changing variables in eq.~(\ref{e2.1.3}), one obtains
\begin{equation}
f_t(r) = \int_{t/(r+1)^{1/\alpha}}^{t/r^{1/\alpha}} L_{\alpha,B}(u) \ \mathrm{d}u,
\end{equation}
which leads to the following asymptotic expression%
\footnote{As $t \simeq S_r = \sum_{i=1}^r \tau_i$ scales as $r^{1/\alpha}$, 
the limit $r \to \infty$ corresponds to the limit $t \to \infty$.}%
:
\begin{equation}
\begin{array}{|c|}
\hline
f_t(r) \mathrel{\mathop{\kern 0pt\longrightarrow}\limits_{t \to \infty}}
\frac{t}{\alpha r^{1+1/\alpha}} \ L_{\alpha,B} \left( \frac{t}{r^{1/\alpha}} \right)
\\
\hline
\end{array}
\label{e2.3.6}
\end{equation}
In sharp contrast with the finite moments case, the renewal number distribution
$f_t(r)$ differs strongly from the distribution of the sum (eq.~(\ref{e2.3.2}))
even though both distributions involve L\'evy stable laws. As will be shown in
\S\ref{s3.1}, $f_t(r)$ has a slow decay at small $r$ and a fast decay 
at large $r$ unlike $\psi^{r*}(t)$ which decays as $1/t^{1+\alpha}$. 

The mean number of renewals $\langle r \rangle_t$ is finite and can be 
related to a negative moment of a L\'evy stable law:
\begin{equation}
\langle r \rangle_t \simeq \int_0^\infty r f_t(r) \ \mathrm{d}r
= t^\alpha \int_0^\infty u^{-\alpha} L_{\alpha, B}(u) \ \mathrm{d}u.
\end{equation}
Hence, using $\int_0^\infty u^{-\alpha} L_{\alpha, B}(u) \ \mathrm{d}u =
1/(B\alpha\Gamma(\alpha))$ and eq.~(\ref{e2.3.4}), one has finally
\begin{equation}
\langle r \rangle_t \mathrel{\mathop{\kern 0pt\longrightarrow}\limits_{t \to \infty}} \frac{\sin(\pi \alpha)}{\pi \alpha} \ 
\left(\frac{t}{\tau_0}\right)^\alpha.
\label{e2.3.m}
\end{equation}
Finally, using eq.~(\ref{e2.1.5}), the renewal density tends asymptotically
to an ever decreasing power law:
\begin{equation}
R(t) \mathrel{\mathop{\kern 0pt\longrightarrow}\limits_{t \to \infty}}
\frac{\sin(\pi \alpha)}{\pi} \
\frac{1}{\tau_0^\alpha t ^{1-\alpha}}.
\label{e2.3.9}
\end{equation}

Physical comments and an example of a renewal process with infinite 
mean waiting time will be presented in \S\ref{s3.1}.

\section{Application to fractal time random walk and subrecoil cooling}
\label{s3}

We have seen in \S\ref{s1.3} that the most important quantities appearing
in biased random walks and in subrecoil cooling are directly related 
to renewal theory: renewal number distribution $f_t(r)$ (position distribution
in biased random walks), mean number of renewals $\langle r \rangle_t$
(peak height in subrecoil cooling) and renewal density $R(t)$
(current in the biased
random walk, momentum distribution in subrecoil cooling).

In this section, we thus apply the results on renewal processes obtained in 
\S\ref{s2} to biased random walks and subrecoil cooling.
We emphasize the physics
consequences of the divergence of the mean waiting time in renewal processes
by opposing, for each application, one example with finite mean waiting time 
and one example with infinite mean waiting time. These examples reveal the
generic features of the cases with finite or infinite mean waiting times.

\subsection{Biased random walks}
\label{s3.1}

\begin{figure}[ht]
\includegraphics[scale=0.33,angle=-90]{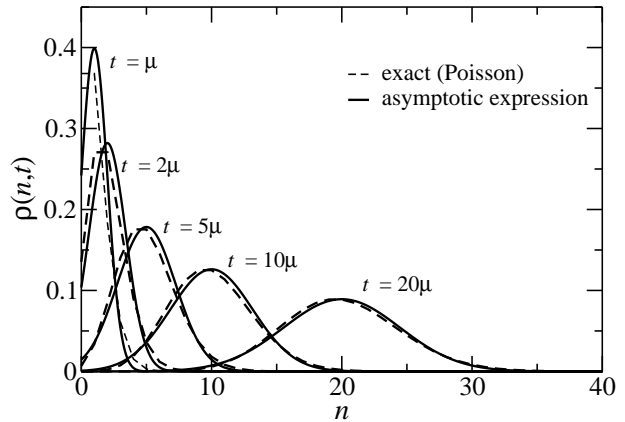}
\caption{Time evolution of the position distribution $\rho(n,t)$
for a biased random walk with an exponential waiting time distribution
({\em finite $\langle \tau \rangle$}). The distribution propagates
at constant speed $1/\mu$ and spreads as $\sqrt{t}$. The exact result 
(eq.~(\ref{e3.1.3})) and 
the asymptotic one (eq.~(\ref{e3.1.4})) agree even at short times.}
\label{fig3_1}
\end{figure}

As an example of waiting time distribution with {\it finite mean and
standard deviation} generating a (non fractal time) random walk, 
we consider an exponential distribution
\begin{equation}
\psi(\tau) = \frac{1}{\mu} \ e^{-\tau/\mu} \quad \mathrm{for}
\quad \tau \geq 0
\end{equation}
with mean and standard deviation both equal to $\mu$.
In this case, the convolutions of $\psi(\tau)$ have the simple explicit
forms of the Gamma (or Erlang) distributions
\begin{equation}
\psi^{n*}(\tau) = \frac{\tau^{n-1}}{(n-1)! \mu^n} \ e^{-\tau/\mu}.
\end{equation}
Thus, applying eq.~(\ref{e2.1.3}), the renewal density $f_t(r)$ and
hence the position distribution $\rho(n,t)$ are exactly known:
\begin{equation}
\rho(n,t) = \frac{e^{-t/\mu}}{n!} \ \left( \frac{t}{\mu}\right)^n. 
\label{e3.1.3}
\end{equation}
(One recognizes the Poisson distribution of mean $\langle r \rangle_t = t/\mu$,
as expected: in this case, the renewal process is a Poisson process.) 
As shown in Fig.~\ref{fig3_1}, this exact position distribution rapidly 
tends to the Gaussian distribution given by eq.~(\ref{e2.2.3}): 
\begin{equation}
\rho(n,t) \mathrel{\mathop{\kern 0pt\longrightarrow}\limits_{t\to \infty}}
 \frac{1}{\sqrt{2\pi t/\mu}} \exp \left[ -
  \frac{(n-t/\mu)^2}{2 t / \mu}\right] \ .
\label{e3.1.4}
\end{equation}

We thus recover
the intuitive picture of normal transport: a position distribution
propagating at a constant speed $1/\mu$ and spreading as $\sqrt{t}$.
The current $i(t)$ resulting from such a distribution in photoconductivity
experiments is  related to the renewal density (eq.~(\ref{e1.3.2})) 
which leads, using eq.~(\ref{e2.2.5}), to
\begin{equation}
i(t) \propto 1/\mu.
\end{equation}

\medskip

\begin{figure}[ht]
\includegraphics[scale=0.33,angle=-90]{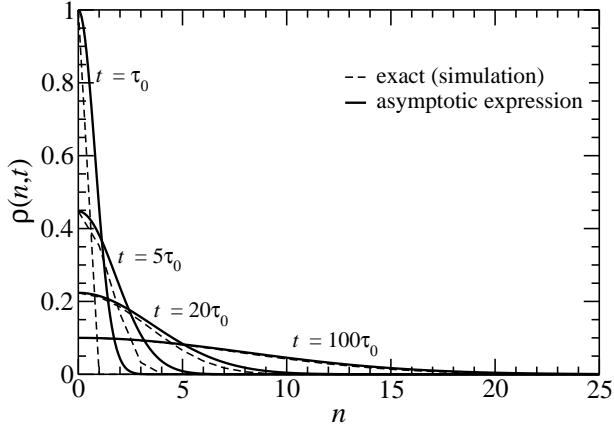}
\caption{Time evolution of the position distribution $\rho(n,t)$
for a biased fractal time random walk constructed from a Pareto distribution of
index $\alpha = 1/2$ ({\em infinite $\langle \tau \rangle$}). 
The distribution spreads slowly towards positive values
but its maximum always remains at the origin. Note the smaller position scale 
and the longer time scales compared to Fig.~\ref{fig3_1}, which emphasizes the
slowness of transport in the fractal time random walk.
The exact (simulated) and asymptotic results (eq.~(\ref{e3.1.7})) are in 
good agreement. (They have seemingly different norms at short times because
the exact result is a discrete distribution while the asymptotic one is a 
continuous distribution.)
}
\label{fig3_2}
\end{figure}

As an example of waiting time distribution with {\it infinite mean and
standard deviation} generating a fractal time random walk, 
we consider a Pareto distribution %
\footnote{For a Pareto distribution of index $\alpha>2$ (finite standard
deviation), the normal transport case is recovered.}
of index $\alpha = 1/2$,
\begin{equation}
\psi(\tau) = \frac{\tau_0^{1/2}}{2 \tau^{3/2}} 
\quad \mathrm{for} \quad \tau \geq \tau_0
\end{equation}
with $\tau_0 >0$. As there is no simple analytic form for $\psi^{n*}(t)$ 
and thus for the position distribution $\rho(n,t) = \Psi^{n*}(t) 
- \Psi^{(n+1)*}(t)$ (eq.~(\ref{e2.1.3})), 
we obtain $\rho(n,t)$ through numerical simulation (see Fig.~\ref{fig3_2}). 
On the other hand, using eq.~(\ref{e2.3.6}) and the known form for the 
asymmetric L\'evy stable law of index $1/2$ called the Smirnov law 
(or using eq.~(\ref{e2.3.3})), the asymptotic distribution is found to 
be a half-Gaussian 
\begin{equation}
\rho(n,t) 
\mathrel{\mathop{\kern 0pt =}\limits_{t\to \infty}}
 \sqrt{\frac{\tau_0}{t}} \ \exp \left( -\frac{\pi \tau_0}{4t} 
n^2\right).
\label{e3.1.7}
\end{equation}
The fact that a half-Gaussian is obtained should {\it not} give the impression
that fractal time transport is similar to normal transport described
by full Gaussians (eq.~(\ref{e3.1.4})). First, the half-Gaussian is specific
to Pareto waiting time distributions with $\alpha = 1/2$ (see below for other 
$\alpha$'s). Second, the properties of the half-Gaussian in 
fractal time transport are completely different from those 
of the full Gaussian of the normal 
transport. Indeed, instead of propagating, the distribution peak remains at the
origin $n=0$ at all times. 
Only the tails spread to the right, more and more slowly
as times goes by. This is due to the fact that the carriers
statistically tend to be trapped into deeper and deeper traps at long times,
which slows down their motion. The resulting current is thus a decreasing 
function of time, at all times. Using eq.~(\ref{e1.3.2}) and eq.~(\ref{e2.3.9}),
one obtains:
\begin{equation}
i(t) \sim \frac{1}{\sqrt{\tau_0 t}}.
\end{equation}

It is worth examining the general case of waiting time distributions 
with power law tails like eq.~(\ref{e2.3.1}) and infinite means ($\alpha <1$).
From the following expansion of asymmetric L\'evy stable laws with $\alpha <1$,
\begin{equation}
L_{\alpha, B}(x) \mathrel{\mathop{\kern 0pt\simeq}\limits_{x\to \infty}}
\frac{\alpha B}{\Gamma(1-\alpha) x^{1+\alpha}},
\end{equation}
one finds the behaviour close to the origin (for $t \to \infty$):
\begin{equation}
\rho(n,t) \mathrel{\mathop{\kern 0pt\simeq}\limits_{n\to 0}} 
	\left( \frac{\tau_0}{t} \right)^\alpha.
\end{equation}
Hence, the position distribution is flat ($n$ independent) at small
$n$, as already observed for the special case $\alpha = 1/2$.
Moreover, using
\begin{equation}
L_{\alpha, B}(x) \mathrel{\mathop{\kern 0pt\simeq}\limits_{x\to 0}}
A x^{\frac{\alpha-2}{2(1-\alpha)}} \exp \left[ -\frac{C}{x^{\alpha/(1-\alpha)}} \right],
\end{equation}
where $A$ and $C$ are constants, one obtains
\begin{equation}
\rho(n,t) \propto \exp\left(- \frac{Cr^{1/(1-\alpha)}}{t^{\alpha/(1-\alpha)}} 
\right),
\end{equation}
up to power law corrections. As $1/(1-\alpha)>1$, the transport front 
always decreases faster than exponentially and presents a well defined
characteristic position. Consequently, the mean carrier position yielding the
current, is well defined, unlike the mean waiting time which is infinite. 
Using eq.~(\ref{e1.3.2}) and eq.~(\ref{e2.3.9}), 
one finds that the current decays as a power law:
\begin{equation}
i(t) \propto \frac{1}{t^{1-\alpha}},
\end{equation}
as already observed for the special case $\alpha = 1/2$ and in agreement with
photoconductivity transient experiments.

\subsection{Subrecoil cooling}
\label{s3.2}

Consider first the case of waiting time distributions with $\alpha = 1/\beta>2$
(eq.~(\ref{e1.2.5})) ensuring a finite mean waiting time 
$\langle \tau \rangle$ (and a finite $\langle \tau^2 \rangle$).
According to eq.~(\ref{e2.2.5}) and eq.~(\ref{e1.2.7}), one has
\begin{equation}
R(t) \mathrel{\mathop{\kern 0pt\longrightarrow}\limits_{t\to \infty}}
\frac{1-\beta}{\tau_0} \left( \frac{p_\mathrm{trap}}{p_0}\right)^\beta
\end{equation}
and thus, applying eq.~(\ref{e1.3.3}) with eq.~(\ref{e1.2.2}) and 
eq.~(\ref{e1.3.4}), one finds the momentum distribution
\begin{equation}
\rho(p,t) = \frac{1- \beta}{2 p_\mathrm{trap}} 
\left[1- e^{- \frac{t}{\tau_0} \left( \frac{|p|}{p_0}\right)^\beta} \right] 
\left( \frac{p_\mathrm{trap}}{|p|}\right)^\beta.
\end{equation}
This distribution has stationary tails:
\begin{equation}
\rho(p,t) \simeq 
\frac{1- \beta}{2 p_\mathrm{trap}} 
\left( \frac{p_\mathrm{trap}}{|p|}\right)^\beta
\quad \mathrm{for} \quad |p| > p_0 \left( \frac{\tau_0}{t} \right)^{1/\beta} 
\end{equation}
but a non stationary peak that increases linearly in time:
\begin{equation}
\rho(p,t) \simeq \frac{1-\beta}{2 p_0^\beta p_\mathrm{trap}^{1-\beta}}\ 
\frac{t}{\tau_0}
\quad \mathrm{for} \quad |p| < p_0 \left( \frac{\tau_0}{t} \right)^{1/\beta}.
\end{equation}
This peak is also obtained directly using relation~(\ref{e1.3.6}) 
between the height $h(t)$
and the mean number of renewals of eq.~(\ref{e2.2.4}).

\begin{figure}[ht]
\includegraphics[scale=0.33,angle=-90]{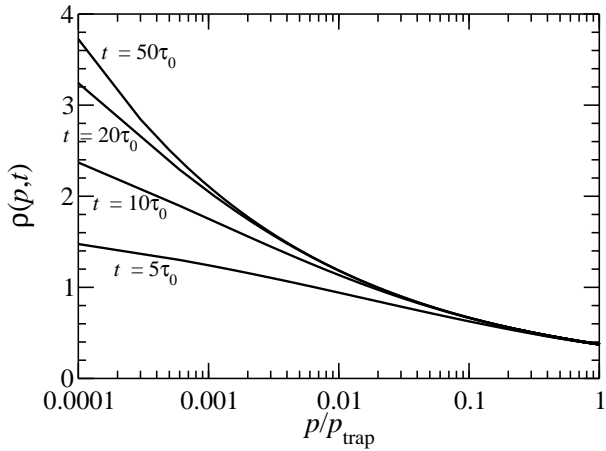}
\caption{Time evolution of the momentum distribution $\rho(p,t)$
for subrecoil cooling with a {\em finite} mean waiting time. Parameters: 
$\beta = 0.25$, $p_0=1$. 
The tails reach a stationary state. Only a vanishingly narrow 
part of the peak goes on increasing at long times (note the logarithmic $p$
scale).
}
\label{fig3_3}
\end{figure}

Finite mean waiting times are not very favourable for the cooling since 
only a vanishingly small fraction of atoms goes on 
accumulating at smaller and smaller velocities (see Fig.~\ref{fig3_3}).

\begin{figure}[ht]
\includegraphics[scale=0.33,angle=-90]{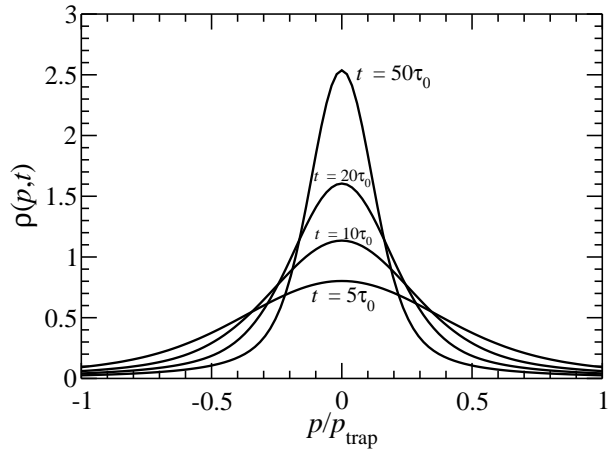}
\caption{Time evolution of the momentum distribution $\rho(p,t)$
for subrecoil cooling with an {\em infinite} mean waiting time. Parameters:
$\beta = 2$, $p_0=1$. The full momentum distribution, tails and peak,
never reaches a stationary state. All atoms go on accumulating 
without limit in a narrower and narrower peak as time increases.
}
\label{fig3_4}
\end{figure}

\medskip
Consider now the case of waiting time distributions with infinite
mean waiting times ($\alpha = 1/\beta <1$) ensuring. According to
eq.~(\ref{e2.3.9}), one has
\begin{equation}
R(t) \mathrel{\mathop{\kern 0pt\longrightarrow}\limits_{t \to \infty}}
\frac{\sin(\pi \alpha) p_\mathrm{trap}}{\pi \alpha \Gamma(\alpha) p_0} \
\frac{1}{\tau_0^\alpha t ^{1-\alpha}}
\end{equation}
(one must write $\psi(t)$ of eq.~(\ref{e1.2.5}) in the form
of eq.~(\ref{e2.3.1}), thus replacing $\tau_0^\alpha$ by 
$\alpha \Gamma(\alpha) (p_0/p_\mathrm{trap}) \tau_0^\alpha$ 
in eq.~(\ref{e2.3.9})).
Thus, applying eq.~(\ref{e1.3.3}) with eq.~(\ref{e1.2.2}) and
eq.~(\ref{e1.3.4}), the momentum distribution writes
\begin{equation}
\rho(p,t) \mathrel{\mathop{\kern 0pt\longrightarrow}\limits_{t \to \infty}}
\frac{\sin(\pi \alpha)}{2\pi \alpha^2 \Gamma(\alpha) p_0}
\ \left( \frac{t}{\tau_0}\right)^\alpha
\mathcal{G}\left( \frac{tp^\beta}{\tau_0 p_0^\beta} \right),
\end{equation}
where 
\begin{equation}
\mathcal{G}(q) = \alpha\int_0^1 \mathrm{d} u \ u^{\alpha-1} e^{-(1-u)q}
\end{equation}
is a confluent hypergeometric function.
Thus $\rho(p,t)$ presents a scaling form and evolves at all times
scales, with no stationary state. The tails behave as 
\begin{equation}
\rho(p,t) \mathrel{\mathop{\kern 0pt\longrightarrow}\limits_{t \to \infty}}
\frac{\sin(\pi \alpha)}{2\pi \alpha \Gamma(\alpha)}
\left( \frac{\tau_0}{t}\right)^{1-\alpha}
\frac{1}{p_0^{1-\beta}p^\beta}
\quad \mathrm{for} \quad |p| > p_0 \left( \frac{\tau_0}{t} \right)^{1/\beta}
\end{equation}
and the peak, also obtained directly from eq.~(\ref{e1.3.6}) 
and the mean number of renewals (\ref{e2.3.m}), behaves as
\begin{equation}
\rho(p,t) \mathrel{\mathop{\kern 0pt\longrightarrow}\limits_{t \to \infty}}
\frac{\sin(\pi \alpha)}{2\pi \alpha^2 \Gamma(\alpha) p_0}
\ \left( \frac{t}{\tau_0}\right)^\alpha
\quad \mathrm{for} \quad |p| < p_0 \left( \frac{\tau_0}{t} \right)^{1/\beta}
\end{equation}
Infinite mean waiting times are favourable for the cooling 
(Fig.~\ref{fig3_4}): all atoms
accumulate in a narrower and narrower peak in the vicinity of $p=0$.
The cooling goes on without fundamental limits at long times. 
The absence of limits is related to the significant weight 
({\it cf. $\langle \tau \rangle = \infty$})
of $p$ states with arbitrarily long lifetimes.

\section{Bibliographical notes}
\label{s4}

\paragraph{Fractal time}
The theory of fractal time random walks was developed 
in particular cases in \cite{MoS1973} using continuous time random walks
and, in the general case, in \cite{Shl1974}, adding Tauberian techniques
to handle Laplace transforms. Fractal times were used in \cite{SWK1987}
to model turbulent diffusion. Several other applications 
involving fractal times were presented in \cite{Shl1988} and \cite{SSB1991}.

\paragraph{Subrecoil cooling}
The statistical approach to subrecoil cooling was developed in three
steps: \cite{BBE1994}, \cite{Bar1995} and \cite{BBA2002}. The connection 
of this approach with renewal theory was stressed in \cite{BaB2000}.

\paragraph{Renewal processes}
The basic theory of renewal processes is presented in a small book 
\cite{Cox1962}. This book does not include infinite mean waiting times but
some general expressions for finite mean waiting times are still
valid in this case too (see \S\ref{s2.1}). Ref.~\cite{Fel1971}, chapter XI, 
presents
renewal theory with a more theoretical viewpoint and includes some
results on infinite mean waiting times which were discovered at the beginning 
of the sixties, in spite of the fact that no application seemed 
to be known at the time.
Renewal processes in a physics context
including infinite mean waiting times
have been studied systematically in \cite{GoL2001} 
using Lapla\-ce transform techniques. 

The mean number of renewals given by eq.~(\ref{e2.3.m})
for infinite mean waiting times agrees with eq. (8) in \cite{Shl1974}
obtained through Tauberian theorems for the fractal time random walk 
(there is to be a misprint in this reference: the expression for 
$\psi(\tau)$, page 424, line 3, must be replaced by $\psi(\tau) \sim
\alpha/\left[ t^{1+\alpha} \Gamma(1+\alpha) A(t) \right]$). It also agrees 
with the renewal theory developed in \cite{GoL2001} (eq.~(3.6)).

The renewal number distribution expressed with L\'evy stable laws 
given by eq.~(\ref{e2.3.6}) agrees with
eq.~(5.6) of \cite{Fel1971}, which is however more complicated.

\paragraph{Generalized central limit theorem}
The generalized central limit theorem and results on L\'evy stable
laws used in \S\ref{s2.3} can be found, {\it e.g.} in Appendix B 
of \cite{BoG1990} or in \S4.2 of \cite{BBA2002}. Details on L\'evy stable
laws can also be found in \S1.2 of \cite{SaT1994}.

\section*{Conclusion}
\label{s5}

This paper has underlined the common statistical core at work 
in two seemingly opposite problems: a diffusion mechanism (fractal time
random walk) and a cooling mechanism (subrecoil laser cooling). This core 
is made of waiting time distributions with infinite means.

Usual theoretical techniques for these problems are based on Tauberian theorems
which, through the Laplace transform, imply some loss of physical intuition.
Here, we have presented a renewal theory approach which, thanks to the 
generalized central limit theorem, provides a shortcut to obtain physically
relevant quantities. It emphasizes the key contribution of the unusual
behaviour of sums of random variables with infinite means.

Understanding these stochastic processes with infinite means is not 
a purely academic game. It has already led to significant improvements
of laser cooling strategies \cite{RBB1995} and other improvements are under 
way \cite{Bar2004}.

\begin{acknowledgments}
I thank O.E. Barndorff-Nielsen, M. Romeo and M. Shlesinger for discussions.
\end{acknowledgments}

\bibliographystyle{apsrev}
\bibliography{bibFB.bib}

\end{document}